\documentclass[11pt,onecolumn,letterpaper]{IEEEtran}
\usepackage[utf8]{inputenc}
\usepackage{amssymb}
\usepackage{graphicx}
\usepackage{amssymb}
\usepackage{epsfig}
\usepackage{balance}
\usepackage{url}
\usepackage{rotating}
\usepackage{listings}
\usepackage{url}
\usepackage{color}
\usepackage{cite}
\usepackage{comment}
\usepackage{tabularx}
\usepackage{enumerate}
\usepackage{longtable}
\usepackage{threeparttable}
\usepackage{setspace}
\usepackage{float}
\usepackage{verbatim}
\usepackage{algorithmic}
\usepackage{array}
\usepackage{amsmath}
\usepackage[tight,footnotesize]{subfigure}
\usepackage{fancyvrb}
\usepackage{lineno}
\usepackage{wrapfig}

\newcommand{\clr}{black}

\makeatletter
\newif\if@restonecol
\makeatother

\usepackage[linesnumbered]{algorithm2e}
\usepackage{verbatim}

\begin{document}

\title{CUDA Leaks:\\ Information Leakage in GPU Architectures }

\author{
	\IEEEauthorblockN{Roberto Di Pietro\IEEEauthorrefmark{1} Flavio Lombardi\IEEEauthorrefmark{1} Antonio Villani\IEEEauthorrefmark{1}}\\
	\IEEEauthorblockA{\IEEEauthorrefmark{1} Department of Maths and Physics\\
					 Roma Tre University\\
					 Rome, Italy\\
					 Email: $\{$dipietro,lombardi,villani$\}$@mat.uniroma3.it}
}

\maketitle
\IEEEdisplaynotcompsoctitleabstractindextext

\IEEEpeerreviewmaketitle

\begin{abstract}
Graphics Processing Units (GPUs) are deployed on most present server, desktop, and even mobile platforms.
Nowadays, a growing number of applications leverage the high parallelism offered by this architecture
to speed-up general purpose computation. This phenomenon is called GPGPU computing (General Purpose GPU computing).\\
The aim of this work is to discover and highlight security issues related to CUDA, the most widespread platform for GPGPU computing.
In particular, we provide  details and proofs-of-concept about a novel set of vulnerabilities CUDA architectures are subject to, 
that could be exploited to cause severe information leak.  
Following (detailed) intuitions rooted on sound engineering security,
we performed several experiments targeting the last two generations of CUDA devices: Fermi and Kepler.
We discovered that these two families do suffer from information leakage vulnerabilities. 
In particular, some vulnerabilities are shared between the two architectures, while others are idiosyncratic of 
the Kepler architecture. 
\begin{color}{\clr}
As a case study, we report the impact of one of these vulnerabilities on a GPU implementation of the AES encryption algorithm.
\end{color}
We also suggest software patches and alternative approaches to tackle the presented vulnerabilities.
To the best of our knowledge this is the first work showing that information leakage in CUDA is possible using just standard CUDA instructions.
We expect our work to pave the way for further research in the field.\footnote{
This paper has been subject to a review process. The timeline is reported below:\\ 
2012-11-04: Submission to Transactions on Information Forensics And Security\\
2013-01-16: Decision to REVISE and RESUBMIT (T-IFS-03001-2012)\\
2013-02-27: Revised Manuscript Submitted (T-IFS-03001-2012.R1)\\
2013-05-28: REJECT AND RESUBMIT (RR) AS A REGULAR PAPER (T-IFS-03001-2012.R1)\\
2013-07-05: Withdrawn from IEEE TIFS, enhanced and submitted to Another Top Notch Transaction}

\end{abstract}


\section{Introduction}

Graphics Processing Units (GPUs) are a widespread and still underutilized resource.
They are available on most present Desktop PCs, laptops, servers and even mobile phones and tablets.
They are often used as cost-effective High Performance Computing (HPC) resources, as in computing clusters \cite{titan}. 

CUDA (Compute Unified Device Architecture - NVIDIA\texttrademark) is by far the most widespread GPU platform, OpenCL (by AMD\texttrademark) being its only competitor.
Most present applications leverage CUDA for speeding-up scientific computing-intensive tasks
or for computational finance operations \cite{computationalfinancegpu}.
GPU computational power is also employed to offload the CPU from  security sensitive computations.
As an example, various cryptographic algorithms have been ported to GPUs 
\cite{aes-gpu, aescuda,aes-xts, symmetriccudageneric, designingGPUencryption, aescudaanalysis, gpuparallcrypto}.
Such applications require the encryption key or other sensitive data
to be present on the GPU device where they are potentially exposed to unauthorized access.
Any kind of information leakage from such applications would seriously hurt the success of the
shared-GPU computing model, where the term shared-GPU indicates all those scenarios where the GPU resource is actually shared among different users,
whether it is on a local server, on a cluster machine or on a GPU cloud \cite{Georgescu:2011:GAC:2082156.2082161} 
(originating the GPU-as-a-Service ---a specialization of the more general class Computing-as-a-Service).

GPUs are increasingly deployed as Computing-as-a-Service on the Cloud \cite{actualgpuinthecloud,vgpuvirtualization}.
In fact, sharing GPU resources brings several benefits such as sparing the cost involved in building  and maintaining a HW/SW GPU infrastructure.
However, the security implications on both GPU computing clusters and on remote GPU-as-a-Service offerings, 
such as those by companies like Softlayers and Amazon, can be dramatic.
Furthermore, in view of the GPU virtualization approach offered by the upcoming NVIDIA VGX Hypervisor \cite{vgxhypervisor},
information leakage risks would dramatically increase. 
In addition, performance-oriented approaches such as CUDADMA \cite{cudadma} cannot but worsen the information leakage problem.
Unfortunately, there is no effective control on how parallel code (a.k.a. kernels) is actually executed on a GPU,
given that CUDA drivers are based on proprietary code.
GPU architecture and hardware/software implementations are not mature enough when it comes to security considerations \cite{nvidiavuln1,nvidiavuln2}.
Indeed, current GPU device drivers are aimed at performance rather than security and isolation, even if 
open source software stacks are beginning to appear from reverse-engineered CUDA device drivers \cite{gdev};
however, functionality, compatibility, and performance are still very unsatisfactory.
As such, it is not possible to verify the internals of GPU middleware and driver executed code.
In addition, OS-level support for GPU computing is in its infancy \cite{kgpu}. 
The vulnerabilities and deficiencies of such closed-source systems can only be discovered
by leveraging both the disclosed details and the black-box runtime analysis of GPUs.

Further, confidentiality over GPU data and computation has to be guaranteed in order for the shared-GPU approach to achieve widespread acceptance.
The aim of this work is to investigate the security of the CUDA 
platform, with a special focus on information leakage. 
This work focused on Linux, as it is the most widely deployed architecture in GPU clusters and clouds \cite{linuxinthecloud}.

\subsection{Contribution}

This paper provides a number of contributions to the problem of secure computing on Graphics Processing Units, with a focus on 
the CUDA GPGPU computing context \cite{gpgpu}.
In particular, following a detailed analysis of the CUDA architecture, source code was designed, built, and instrumented in order to
assess the security of the  CUDA architecture. 
It is worth noting that we leveraged perfectly standard GPU code to show the information leakage flaws.
As for the first vulnerability described in this work, we were able to induce information leakage on GPU shared memory.
Further, an information leakage vulnerability based on GPU global memory,
and another one based on GPU register spilling over global memory were discovered.
\begin{color}{\clr}
As a case study we evaluated the impact of one of these leakages on a publicly available GPU implementation of cryptographic algorithms. 
In particular, we demonstrated that through the global memory vulnerability it is possible to access both the plaintext and the encryption key. 
\end{color}
The interesting results we obtained and described in this paper open a broad new area. 
The discovered vulnerabilities could 
be used to successfully attack commercial security-sensitive algorithms such as the encryption algorithms running on GPUs.\\
Finally, this paper proposes and discusses countermeasures and alternative approaches to fix the presented information leakage flaws.

\subsection{Roadmap}
This work is organized as follows:
Section \ref{cuda} discusses publicly available details of the CUDA architecture.
Section \ref{ourapproach} describes the rationales behind the discovered vulnerabilities and provides attack details.
Section \ref{experimental} introduces the experimental setup and attacks implementation and shows experimental results.
\begin{color}{\clr}
Section \ref{aes-leakage-glb} describes in detail a case study based on AES CUDA implementation and analyzes the results.
\end{color}
Section \ref{remedies} discusses possible remedies. 
Section \ref{related} introduces related work on CUDA and state-of-the-art results on information leakage.
Finally, Section \ref{conclusion} provides some final considerations and directions for future work. 

\section{CUDA Architecture}
\label{cuda}
CUDA is a parallel computing platform for NVIDIA GPUs. 
CUDA represents the latest result of GPU evolution: old GPUs supported only specific fixed-function pipelines, whereas recent GPUs are 
increasingly flexible.
In fact, General Purpose GPU computing (GPGPU) \cite{gpgpu2} allows deploying massively parallel computing on COTS hardware
where the GPU can be used as a ``streaming co-processor''. 
In this context, present SDKs allow to simplify access to GPU resources to solve a broad set of problems \cite{gpgpu}. 

\begin{figure}[t]
 \centering
 \includegraphics[scale = 0.3]{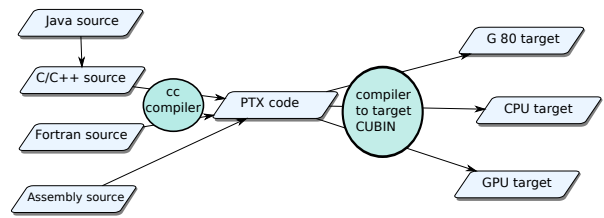}
 \caption{Main steps in compiling source code for CUDA devices}
 \label{fig:CUDA_Compiling}
\end{figure}

The CUDA Software Development Kit (SDK) introduces concepts such as \emph{stream}s and \emph{kernels}.
\emph{kernels} are special functions, that, when called, are run $N$ times in parallel by $N$ different CUDA threads. 
The SDK allows the programmer to write kernel and host code using an enhanced version of the traditional C, C++ and FORTRAN programming languages (as well as Java see \cite{jcuda}).
The number of CUDA threads that run a kernel in parallel for a given call is specified at runtime through the \emph{execution configuration}. 
By changing configuration parameters it is possible to group threads in one or more blocks, depending on the parallel task that must be performed.

CUDA source code is thus split into two components, host code (run by the CPU) and device code (run by the GPU).
This latter will get compiled to an intermediate language called PTX \cite{cudadevguide}.
There exist several PTX versions: one for each \textit{compute capability} (NVIDIA way of defining hardware versions, see Table \ref{tab:cudacomputecapabilities}).
In addition, at kernel deployment/install time, such intermediate code is further compiled by the CUDA device driver to binary CUBIN code,
which is actually tailored to the specific target physical GPU where it will be executed.
This approach allows specific code optimization to be tied to the actual GPU architecture. 
The CUDA compilation and deployment process is depicted in Figure \ref{fig:CUDA_Compiling}.

The CUDA architecture can be synthesized  as follows:
\begin{itemize}
 \item a binary file comprising host and PTX \cite{cudadevguide} object code;
 \item the CUDA user-space closed source library (libcuda.so);
 \item the NVIDIA kernel-space closed source GPU driver (nvidia.ko);
 \item the hardware GPU with its interconnecting bus (PCI Express or PCIe),
 memory (Global, Shared, Local, Registers) and computing cores (organized in Blocks and Threads)---(see Figure \ref{fig:CUDA_Architecture} \cite{cudacs}).
\end{itemize}

CUDA allows the developer to use two different APIs: runtime API and driver API.
The former frees the programmer from tedious low-level tasks but does not allow specifying finer interaction details with the hardware.
The latter allows accessing some inner ``obscure'' features of the GPU and as such it can be leveraged to obtain better performance
and/or to enable advanced features.
In particular, the driver API is implemented in the nvcuda dynamic library which is copied on the system during the installation of the device driver.
It is a handle-based, imperative API: most objects are referenced by opaque handles that may be specified to functions to manipulate the objects.
A CUDA context is analogous to a CPU process.
Contexts are leveraged by CUDA to handle relevant tasks such as virtual memory management for both host and GPU memory.
All resources and actions performed within the driver API are encapsulated inside a CUDA context,
and the system should automatically clean up these resources when the context is destroyed.
However, this is just a working hypothesis (even if corroborated by experimental results), since implementation details are not known, as already mentioned above.
Applications manage concurrency through CUDA streams.
A stream is a sequence of commands (possibly issued by different host threads) that execute in order.
Different streams may execute their commands in any order with respect to one another---this form of parallelism comes with no guarantee on streams scheduling. 

\begin{center}
\begin{table*}
\caption{CUDA \textit{Compute Capability} (CC) version, introduced novelty, and potential sources of security issues}
\resizebox{\columnwidth}{!}{
\begin{tabular}{|l|l|l|}
\hline 
\textit{CC} & Novelty w.r.to previous version & Potential Leakage due to\\\hline \hline
1.0 & original architecture & memory isolation\\\hline
1.1 & atomic op. on global mem. & memory isolation\\\hline
1.2 & atomic op. on shared mem., 64-bit words, \textit{warp}-vote functions & memory isolation\\\hline
1.3 & double precision floating point &  memory isolation \\\hline
2.0 & 64-bit addressing, Unified Virtual Addressing, GPUDirect & memory isolation, GPUDirect issues\\\hline
2.1 & performance improvements& memory isolation, GPUDirect issues\\\hline
3.0 & enhanced stream parallelism and resource sharing & memory isolation, GPUDirect issues\\\hline
3.5 & Dynamic Parallelism, Hyper-Q & memory isolation, GPUDirect issues,\\
    &  & Hyper-Q isolation issues\\\hline
\end{tabular}
}
\label{tab:cudacomputecapabilities}
\end{table*}
\end{center}
\normalsize

\subsection{CUDA Memory Hierarchies}

\begin{wrapfigure}{r}{0.5\textwidth}
 \centering
 \includegraphics[scale = 0.35]{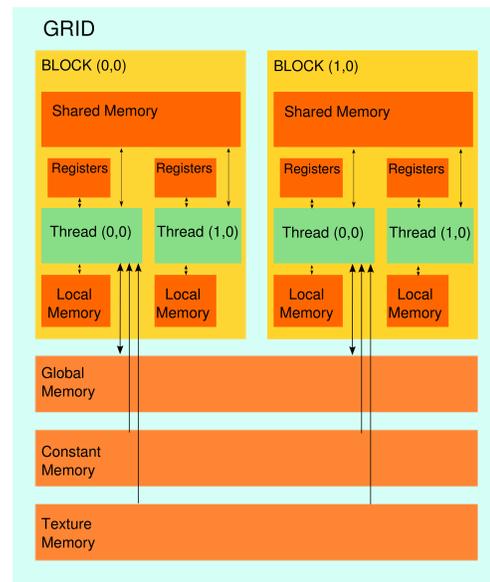}
 \caption{CUDA memory hierarchy and threads}
 \label{fig:CUDA_Architecture}
\end{wrapfigure}

To improve performance, CUDA features several memory spaces and memory types (e.g. global memory, shared memory).
Main memory hierarchy layers are depicted in Figure \ref{fig:CUDA_Architecture}.
Please note that Figure \ref{fig:CUDA_Architecture} shows only a logical organization
of CUDA memory hierarchy; for instance, depending on the size of the reserved memory,
the compiler may choose to map local memory on registers or global memory.
All threads have access to the same global memory.
Each CUDA thread has private local memory.
Each thread block has shared memory visible to all threads of the block and with the same lifetime as the block.
In fact, what happens to the shared memory when a block completes its execution is not specified.
There are also two additional read-only memory spaces accessible by all threads: the constant and texture memory spaces.
The global, constant, and texture memory spaces are persistent across kernel launches by the same application \cite{cudadevguide}.

\subsubsection{Global Memory}

Global memory is accessed via 32-, 64-, or 128-byte memory transactions. 
This is by far the largest type of memory available inside the GPU. 
When a \textit{warp} (i.e. a group of threads, the minimum size of the data processed in SIMD fashion) executes an instruction that accesses global memory,
it coalesces\cite{coalescing} the memory accesses of the threads within the \textit{warp} into one or more memory transactions. 
This allows reducing memory access latency.

\subsubsection{Shared Memory}

Shared memory is expected to be much faster than global memory. 
Such a memory is located near each processor core in order to obtain low-latency access (similarly to cache memory).
Each multiprocessor is equipped with the same amount of shared memory.
The size of the shared memory is in the order of Kilobytes (e.g. 16KB or 64KB times the number
of the available multiprocessors).
Thanks to shared memory, threads belonging to the same block can efficiently cooperate by seamlessly sharing data.
The information stored inside a shared memory bank can be accessed only by threads belonging to the same block. 
Each block can be scheduled onto one multiprocessor per time. 
As such, a thread can only access the shared memory available to a single multiprocessor.
The CUDA developers guide \cite{cudadevguide} encourages coders to make use of this memory as much as possible.
In particular, specific access patterns are to be followed to reach maximum throughput.
Shared memory is split into equally-sized memory modules, called banks, which can be accessed simultaneously.

\subsubsection{Local Memory}

Local memory accesses usually only occur for some automatic variables.
This kind of memory is usually mapped onto global memory, so accesses have the same high latency
and low bandwidth as global memory and are subject to the same requirements for memory coalescing \cite{coalescing}.
Local memory is organized such that consecutive 32-bit words are accessed by consecutive thread IDs.
Accesses are therefore fully coalesced as long as all threads in a \textit{warp} access the same relative address.
On devices with compute capability  2.x or newer, local memory accesses
are always cached in L1 and L2, similarly to global memory \cite{cudadevguide}.
It is worth noting that the L1 cache size can be set at compile time.
This allows the programmer to fine tune data access latency according to requirements.

\subsubsection{Registers}

CUDA registers represent the fastest and smallest latency memory of GPUs. 
However, as we will show later, CUDA registers are prone to leakage vulnerabilities. 
The number of registers used by a kernel can have a significant impact on the 
number of resident \textit{warps}: the fewer registers a kernel uses, the more 
threads and thread blocks are likely to reside on a multiprocessor, improving performance.
Therefore, the compiler uses heuristics to minimize register usage through \emph{register spilling}.
This mechanism places variables in local memory that could have exceeded the number of available registers

\subsection{Preliminary considerations}

The CUDA programming model assumes that both the host and the device maintain their own \emph{separate memory spaces} in DRAM,
respectively host memory and device memory.
Therefore, a program manages the global, constant, and texture memory spaces through calls to the CUDA runtime.
This includes device memory allocation and deallocation as well as data transfer between host and device memory.
Such primitives implement some form of memory protection that is worth investigating.
In fact, it would be interesting to explore the possibility of accessing these memory areas bypassing such primitives.
Moreover, this would imply to analyze what kind of memory isolation is actually implemented.
In particular, it would be interesting to investigate whether it is possible to obtain a specific global memory location by leveraging GPU allocation primitives.
Further, when two different kernels $A,B$ are being executed on the same GPU,
it would be interesting to know what memory addresses can be accessed by kernel $A$
or if $A$ can read or write locations that have been allocated to $B$ in global memory.
What happens to memory once it is deallocated is undefined, and released memory is not guaranteed to be zeroed \cite{zeroizzazione, zeroizzazione2}.
Finally, the trend in memory hierarchy is to have a single unified address space between GPU and CPU (see also \cite{cudadevguide}).
An example is the GMAC \cite{gmac} user-level library that implements an Asymmetric Distributed Shared Memory model.
In fact, the CUDA unified virtual address space mode (Unified Virtual Addressing) puts both CPU and GPU execution in the same address space.
This alleviates CUDA software from copying data structures between address spaces, 
but it can be an issue on the driver side since GPUs and CPUs can compete over the same memory resources.
In fact, such a unified address space can allow potential information leakage. 

Table \ref{tab:cudacomputecapabilities} summarizes existing \textit{compute capabilities}
and the main novelty with respect to the previous version of the architecture.
The third column shows the potential leakage areas that specific versions of the platform can be subject to,
stemming from its architectural characteristics and functionality.

\section{Rationales of vulnerabilities research}
\label{ourapproach}

The strategy adopted by IT companies to preserve trade secrets consists in not revealing much details
about the internals of their products.
Although this is considered the best strategy from a commercial point-of-view, for what concerns security 
this approach usually leads to unexpected breaches \cite{securitythroughobscurity}.
Despite this serious drawback, the security-through-obscurity approach has been embraced by the graphics technology companies as well.

Many implementation details about the CUDA architecture are not publicly available. 
Once a program invokes a CUDA routine it partially looses the control over its data. 
In particular, uncertainty increases when data is transferred to the GPU. 
If we only consider the public information about the architecture, 
it is unclear if any of the security constraints that are usually enforced in the Operating System (OS), 
are maintained inside the GPU.
The only implementation details available via official sources just focus on performance. 
For instance, NVIDIA describes in details which are the suggested access patterns to global memory in order to achieve the highest throughput.
In contrast, important implementation details about security features are simply omitted.
For instance,  there is no official information about the internals of CUDA memory management: it is undefined/uncertain whether memory is zeroed \cite{zeroizzazione2} after releasing it.

Our working hypothesis as for the strategy adopted by GPU manufacturers is that they lean to trade-off performance against security.
Indeed, one of the main objectives of the GPGPU framework is to speed-up computations in High Performance Computing.
In such a context, the memory initialization after each kernel invocation could introduce a non-negligible overhead \cite{zeroingoverhead}. 

To make things worse, NVIDIA implemented memory isolation between different \emph{cudaContext} within its closed source driver.
Such a choice can introduce vulnerabilities, as explained in the following example.
Suppose that an host process $P_i$ needs to perform some computation on a generic data structure $S$.
The computation on $S$ needs to be offloaded to the GPU for performance reasons.
Hence, $P_i$ allocates some host memory $M_h^i$ to store $S$;
then it reserves memory on the device $M_d^i$ and copies $M_h^i$ to $M_d^i$ using the CUDA runtime primitives.
From this moment onwards, the access control on $M_d^i$ is not managed by the host OS and CPU.
It becomes exclusive responsibility of the NVIDIA GPU driver. 
That is, in this context the driver takes the place of the Operating System.
Due to their importance, the isolation mechanisms provided by an Operating System are usually subject to a thorough review.
The same is not true for GPU drivers.
Hence, this architecture raises questions such as whether it is possible for a process $P_j$ to circumvent
the GPU Virtual Memory Manager (VMM) and obtain unauthorized access to the GPU memory of process $P_i$. 

Providing memory isolation in CUDA is probably far more complex than in traditional architectures.
As described in Section~\ref{cuda}, CUDA threads may access data from multiple memory spaces during their execution.
Although this separation allows to improve the performance of the application,
it also increases the complexity of access control mechanisms and makes it prone to security breaches.
Memory transfers across the PCIe bus for the global, constant, and texture memory spaces are costly.
As such, they are made persistent across kernel launches by the same application.
This implies that the NVIDIA driver stores application-related \emph{state information} in its data structures.
As a matter of fact, in case of interleaved execution of CUDA kernels belonging to different host processes,
the driver should prevent process $P_i$ to perform unauthorized access to memory locations reserved to any process $P_j$.
Indeed, as it will be proved and detailed in Section~\ref{experimental}, this mechanism has a severe flaw and could leak information.

A solution that preserves isolation in memory spaces like global memory,
that in the newer boards reaches the size of several Gigabytes, 
could be unsuitable for more constrained resources like shared memory or registers.
Indeed, both shared memory and registers have peculiarities that rise the level of complexity for the memory isolation process.
The shared memory, for example, is like a cache memory which is directly usable by the developers. 
This is in contrast with more traditional architectures such as \emph{x86} where software is usually cache-oblivious.
For what concerns registers, a feature that could taint memory isolation is that registers can be used to access global memory as well:
in fact, a modern GPU feature (named \emph{register spilling}) allows to map a large number of kernel variables onto a small number of registers.
When the GPU runs out of hardware registers, it can transparently leverage global memory instead.

In order to investigate the aforementioned questions, we performed a few experiments on two different
generations of CUDA-enabled devices (Fermi and Kepler) using a black-box approach.
It is important to note that in our experiments we consider an adversary which is able to interact
with the GPU only through legitimate invocations to CUDA Runtime.

\section{Experimental Results}
\label{experimental}

This section describes the test campaign we set-up on GPU hardware to investigate the possible weaknesses of CUDA architectures.
Following the results of the analysis of the CUDA architecture and according to the rationales for attacks described in the previous section,
a number of experiments were scheduled.
As mentioned in Section~\ref{cuda}, a process in CUDA terminology is identified by its \emph{cudaContext}.
In fact, a host thread  usually gets assigned a single context \emph{cudaContext}.
The performed experiments aim at discovering if, and under what conditions,
a \emph{cudaContext} $C_a$  can maliciously access data belonging to another \emph{cudaContext} $C_b$.
If this happens, it would be a violation of the memory isolation mechanisms.

Tests were performed on the Linux platform using different CUDA HW/SW configurations.
The rationale behind this choice is that the vast majority of shared/distributed GPU computing offerings (GPU clusters and GPU-as-a-Service), are hosted on Linux \cite{linuxinthecloud}.

In the following subsections we will detail three different leakage attacks targeted at different memory spaces.
Each leakage has specific preconditions and characteristics.
As a consequence, proposed attack countermeasures will be quite different.
For each kind of leakage, we developed a C program making use of the standard CUDA Runtime Library.
In just a single case we had to directly write PTX assembly code in order to obtain the desired behavior.

The rest of this section is organized as follows: the experimental testbed is described in detail in Subsection ~\ref{sub:exp_setup};
in Subsection~\ref{sub:shmem_attack} the first and simplest leakage is discussed regarding shared memory;
in Subsection~\ref{sub:glbmem_attack} a potentially much more extended (in size) information leakage is detailed---while not leveraging shared memory---;
finally the most complex and powerful information leakage is described in Subsection~\ref{sub:regmem_attack}.
It leverages registry usage and local memory.

\begin{center}
\begin{table}
	\centering
	\caption{Summary of the results of the experiments}
	\begin{tabular}{| l | l | l |}
		\hline 
		& \textbf{Leakage} & \textbf{Preconditions} \\
		\hline \hline
		\textit{Shared} & Complete & $P_a$ is running \\
		\textit{memory} &  &   \\\hline
		\textit{Global} & Complete  & $P_a$ has terminated and $P_b$  \\
		\textit{memory}	&           & allocates the same amount of\\
				&	    &  memory as $P_a$ \\\hline
		\textit{Registers} & Partial & none \\\hline	 	 	
	\end{tabular}
	\smallskip
	\smallskip
\label{tab:test-summary}
\end{table}
\end{center}

\subsection{Experimental Setup}
\label{sub:exp_setup}

The experimental testbed is composed by COTS CUDA hardware and production-level SDKs.
In order to verify whether the obtained information leakage was independent from the implementation of a specific GPU,
and thus to make our experiments more general, two radically different configurations were chosen.
On the one hand, a Tesla card that can be considered targeting the 
HPC sector;
on the other hand, a GeForce card targeted at consumers and enthusiasts. 
The Tesla card implements the Fermi architecture whereas the GeForce card belongs to the newer Kepler family, i.e. the latest generation of NVIDIA GPUs.
As such, the two GPUs differ with respect to the supported CUDA Capability (2.0 for the Fermi and 3.0 for the GeForce).
In our experiments the compiling process took into consideration the differences between the target architectures.
In Table \ref{tab:testbed} we report the specifications of the two GPUs. 
The reported size of shared memory and registers represent the amount of memory available for a single block (see \cite{cudadevguide}).

Each experiment was replicated on both configurations;
in some cases we tuned some of the parameters to explicitly fit the GPU specifications (e.g. the size of shared memory).

\begin{center}
\begin{table}
	\centering
	\caption{The testbed used for the experiments}
	\begin{tabular}{| l | c | c |}
		\hline
	 	\textbf{GPU Model} & Tesla C2050 & GeForce GT 640 \\
		\hline
		\textbf{CUDA Driver} & 4.2 & 5.0 \\
		\hline
	 	\textbf{CUDA Runtime} & 4.2 & 4.2 \\
	 	\hline
	 	\textbf{CUDA Capability} & 2.0 & 3.0 \\
	 	\hline
	 	\textbf{GPU Architecture} & Fermi GF100 & Kepler GK107 \\
	    \hline
	 	\textbf{Global Memory} & 5 GB & 2 GB \\
	 	\hline
	 	\textbf{Shared Memory per MP} & 48 KB & 16 KB \\
	    \hline
	 	\textbf{Registers} & 32768  & 65536 \\
	 	\hline
	 	\textbf{Warp Size} & 32 & 32 \\ 	
	 	\hline
	 	\textbf{CUDA Cores} & 448 & 384 \\
	 	\hline
	 	\textbf{Multiprocessors (MP)} & 14 & 2 \\
	 	\hline
	 	\textbf{Total Shared Memory} & 672KB & 32KB \\
	 	\hline
		\textbf{Linux Kernel} & 3.3.7-x86\_64 & 3.5.2-x86\_64 \\
	 	\hline
	\end{tabular}
	\smallskip
	\smallskip
\label{tab:testbed}
\end{table}
\end{center}

\subsection{Shared Memory Leakage}
\label{sub:shmem_attack}

\begin{figure}[t]
\begin{center}
	\includegraphics[scale=0.55]{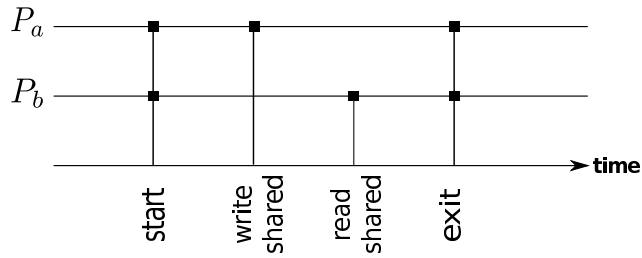}
	\caption{The schedule that causes the leakage on shared memory.
	}
	\label{fig:shmemmem_attack}
\end{center}
\end{figure}

In this scenario the objective of the adversary is to read information stored in shared memory by other processes.
In order to do so the adversary uses regular runtime API functions. 
Present CUDA Runtime allows each \emph{cudaContext} to have exclusive access to the GPU. 
As a consequence, CUDA Runtime does not feature any preemption mechanism\footnote{The only preemption available in CUDA is related to blocks belonging to the same kernel.
In fact, blocks can be preempted during their execution and assigned to another multiprocessor depending on the scheduler}.
The Runtime keeps on accepting requests even when the GPU is busy running a kernel. 
Such requests are in fact queued and later run in a FIFO style.
It is worth noting that the above mentioned requests can belong to different CUDA Contexts.
As such, if no memory cleaning functionality is invoked, when a context-switch occurs, information could be spilled.

\begin{color}{\clr}
\label{leakagequality} 
In fact, every time a malicious process is rescheduled on the GPU, 
it can potentially read the last-written data of the previous process that used the GPU. 
As such, the scheduling order affects which data is exposed.
As an example if $P_a$ is performing subsequent rounds of an algorithm,
the state of the data that can be read reflects the state reached by the algorithm itself.
\end{color}

The experiment to validate such hypothesis is set-up as follows:
two different host-threads belonging to distinct processes are created;
$P_a$ being the honest process, and $P_b$ the malicious one trying to sneak trough $P_a$ memory.
$P_a$ executes $K$ times a kernel that writes in shared memory ($K=50$ in this test).
In this experiment $P_a$ copies a vector $V_g$.
Every element in $V_g$ is of type \emph{uint32\_t} as defined in the header file \emph{stdint.h}.
The size of the vector is set equal to the size of the physical shared memory:
48KB and 16KB respectively for the Tesla C2050 and the GeForce GT 640.
The copy proceeds from global memory to shared memory. 
The host thread initializes $V_g$ deterministically using incremental values (i.e. $V_g[i] = [i]$).
$P_b$ executes $K$ invocations of a kernel that reads shared memory. 
In particular, $P_b$ allocates a vector $V_g$ and declares in shared memory a vector $V_s$ of size equal to the shared memory size.
Then data is copied from $V_s$ to $V_g$.

With this particular sequence of operations 
\begin{color}{\clr}
(see Figure \ref{fig:shmemmem_attack})  
\end{color}
\label{leakagequantity} 
$P_b$ recovers exactly the same set of values written by $P_a$ in shared memory during its execution.
That is, a complete information leakage happens as regards the content of memory used by $P_a$.
One of the parameters of the function developed for this experiment is the kernel block and grid size.
In our tests we have adopted a number of blocks equal to the actual number of physical multiprocessors of the GPU.
As regards the number of threads we have specified a size equal to the \textit{warp}-size.

Quite surprisingly, the values captured by $P_b$ appear exactly in the same order as they were written by $P_a$.
This is not an obvious behavior. 
In fact, given that the GPUs feature a block of shared memory for each multiprocessor, 
a different scheduling of such multiprocessors would lead to different orderings of the read values.
In order to better investigate this issue, we make use of another vector 
where the first thread of each thread-block writes the ID of the processor it is hosted on. 
This info is contained in the special purpose register $smid$ that the CUDA Instruction Set Architecture (ISA) is allowed to read.
In fact, $smid$ is a predefined, read-only special register that returns the processor (SM) identifier on which a particular thread is executing.
The SM identifier ranges from $0$ to $nsmid-1$, where $nsmid$ represents the number of available processors.
As a consequence, in order to read this information, we embedded the following instruction, written in PTX assembler, inside the kernel function:
\begin{Verbatim}[frame=single]
asm("mov.u32 %0, \%smid;" : "=r"(ret) );
\end{Verbatim}
The above instruction copies the unsigned 32-bit value of register \emph{\%smid} into the \emph{ret} variable residing in global memory.
Note that CUDA deterministically chooses the multiprocessors for the first $nsmid$ blocks.
As an example, for the Fermi card we obtained the multiprocessor ID sequence: 0, 4, 8, 12, 2, 6, 10, 13, 1, 5, 9, 3, 7, 11.
This is good news for predictability and for information leakage, 
and it is a valid explanation for the fact that process $P_a$ was able to read the values in the same order they were written by process $P_b$.
In order to obtain the information leakage, it is essential for the kernel of $P_b$ to be scheduled on the GPU before the termination of host process $P_a$.
In fact, we experimentally verified that shared memory is zeroed by the CUDA runtime before the host process invokes the \emph{exit()} function.

\subsection{Global Memory Leakage}
\label{sub:glbmem_attack}
	\begin{figure}[t]
		\begin{center}

		\includegraphics[scale=0.55]{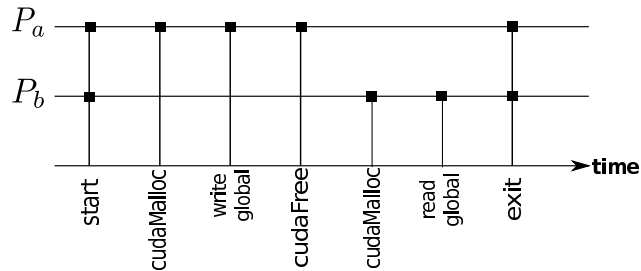}
		\caption{The schedule that causes the leakage on global memory. 
	}
		\label{fig:glbmem_attack}
	\end{center}
\end{figure}

In this scenario the adversary manages to get unauthorized access to information contained in GPU Global Memory.
The vulnerability that is exploited is due to the lack of memory-zeroing operations. 

As before, we have two independent host processes, namely $P_a$ and $P_b$, representing the honest and the malicious process, respectively.
In this experiment \begin{color}{\clr}
(see Figure \ref{fig:glbmem_attack})
\end{color}
 $P_a$ is executed first and allocates four vectors $V_1, V_2, V_3, V_4$ of size $D$ bytes each in GPU memory
(dynamic allocation on the GPU uses the \emph{cudaMalloc()} primitive).
The $i$-th elements of vectors $V_1$ and $V_2$ are initialized in the host thread as follows:

$$ V_1[i] = i ; V_2[i] = D + i $$

We set $D$ equal to 64 KB.
As such, by juxtaposing $V_1$ and $V_2$ we obtain all values from $0$ to $64K - 1$.
Then, $P_a$ invokes a kernel that copies $V_1$ and $V_2$ into $V_3$ and $V_4$, respectively.
$P_a$ then terminates and $P_b$ gets scheduled.
$P_b$ allocates four vectors $V_1, V_2, V_3, V_4$ of size $D$ bytes each, just as $P_a$ did before.

It is worth noting that $P_b$ does not perform any kind of vector initialization.
$P_b$ now runs the very same kernel code $P_a$ executed before and copies $V_3$ and $V_4$ back in the host memory.
At this point, 
$P_b$ obtains exactly the same content written before by $P_a$.
As a consequence, we can say that the information leakage on this kind of memory is full.

By allocating different amounts of memory, partial leakage is obtained,
whereas a total leakage can be obtained only by reallocating exactly the same amount of memory released before.

\subsection{Register-Based Leakage}
\label{sub:regmem_attack}

\begin{figure}[t]
\begin{center}
	\includegraphics[scale=0.55]{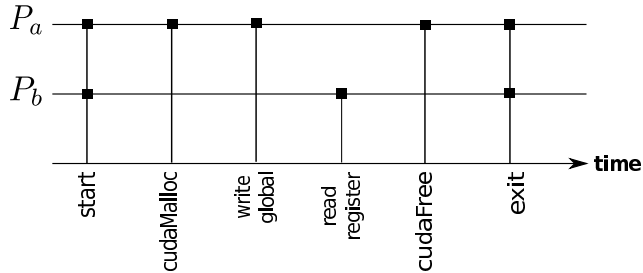}
	\caption{The schedule that causes the leakage on Registers.
			In this scenario $P_b$ accesses the global memory without runtime primitives.}
	\label{fig:regs_attack}
\end{center}
\end{figure}
\begin{figure}
{
	\fontsize{8pt}{10pt}\selectfont
	\begin{verbatim}
	__device__
	void get_reg32bit(uint32_t *regs32) {
	\end{verbatim}
	\begin{quote}
	
		\begin{verbatim}
		# declaration of 8300 registers
		asm(".reg .u32 r<8300>;\n\t");
		# move the content of register r0 into
		# the position 0 of regs32[]
		asm("mov.u32 %0, r0;" : "=r"(regs32[0]));
		asm("mov.u32 %0, r1;" : "=r"(regs32[1]));
		...
		asm("mov.u32 %0, r8191;" : "=r"(regs32[8191]));
		\end{verbatim}
	
	\end{quote}
	\begin{verbatim}
	}
	\end{verbatim}

}
\label{fig:snip-reg-attack}
\caption{A snippet of the code that allows to access global memory without \emph{cudaMalloc}}
\end{figure}

\begin{center}
\begin{table}
	\centering
	\caption{Number of bytes leaked with two rounds of the register spilling exploit. Different
			 rounds could leak different locations of global memory. We do not consider repetitions (i.e.
			 an address that leaks in both rounds)}
	\begin{tabular}{| l | l | l | l | l |}
	\hline
		& \textbf{32KB} & \textbf{64KB} & \textbf{128KB}\\
		 \hline
		 \textbf{16MB} &  32K & 128K & 256K \\
		 \hline
		 \textbf{32MB} &  64K & 128K & 256K \\
		 \hline
		 \textbf{64MB} &  64K & 64K  & 128K \\
		 \hline
		 \textbf{128MB}&  64K & 64K  & 256K \\
	\hline 	 	 	
	\end{tabular}
	\smallskip
	\smallskip
\label{tab:reg-attack-summary}
\end{table}
\end{center}

\begin{figure}
	\centering
	\includegraphics[scale=0.35,angle=-90]{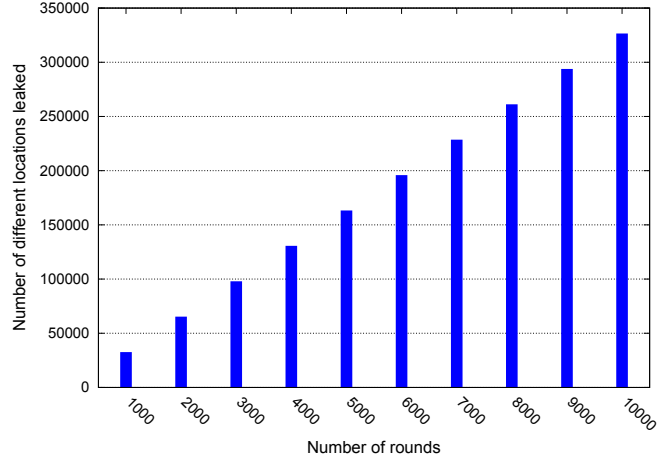}
	\caption{The number of different locations leaked depends on the number of rounds.}
	\label{fig:reg-rounds-kepler}
\end{figure}

In this last described leakage, the adversary makes use of GPU registers to access global memory.
In fact, by using the PTX intermediate language (see Figure \ref{fig:CUDA_Compiling}), 
we are able to exactly specify in a kernel how many registers it will actually need during execution.
If the required number of registers exceeds those physically available on chip,
the compiler (PTX to CUBIN) starts spilling registers and actually using global memory instead.
The number of available registers per each block depends on the chip capability
(i.e. 32K for CUDA capability 2.0 and 64K for CUDA capability 3.0). 

In Figure 6
a PTX code fragment that can be used by the kernel for register reservation is shown.
From the point of view of an adversary, register spilling is an easy way to access global memory bypassing runtime access primitives.
In our experiments we tried to understand whether such an access mechanism to global memory
would undergo the same access controls memory allocation primitives (e.g. \emph{cudaMalloc()}) are subject to.

Surprisingly, we discovered this was not true and we found out the described mechanism could allow a malicious process to access
memory areas that had been reserved to other \emph{cudaContexts}.
It is important to stress that the malicious process can access those locations
even while the legitimate process still owns them (namely before it calls the \emph{cudaFree()}).
This is the reason why we believe this latter leakage is the most dangerous among the presented ones.

Actually, we were able to replicate such an attack exclusively on the Kepler GPU.
However, we are investigating the configuration parameters that cause the information leaks on Fermi. 

In the remaining part of this section we describe in details the steps required to obtain the leakage. 
As a preliminary step, we zeroed the whole memory available on the device in order to avoid tainting the results. 
As for previous experiments, we needed two independent host processes, $P_a$ and $P_b$, as the honest and the malicious processes, respectively.

$P_a$ performs several writes into the global memory whereas $P_b$ tries to circumvent the memory isolation mechanisms to read those pieces of  information. 
In order to verify in a more reliable way the attack outcome, $P_a$ writes a pattern that is easy to recognize;
in particular $P_a$ allocates an array and marks the first location with the hexadecimal value \emph{0xdeadbeef}.
At position $j$ of the array, $P_a$ stores the value \emph{0xdeadbeef} + $j$ where $j$ represents 
the offset in the array represented in hexadecimal.

$P_b$ reserves a predefined number of registers and copies the content of the registers back to the host memory;
in this way $P_b$ tries to exploit the register spilling to violate the memory locations reserved by $P_a$. 
Indeed, if the register spilling mechanism is not properly implemented, 
then some memory locations reserved to $P_a$ could be inadvertently assigned to $P_b$. 
We ran $P_a$ and $P_b$ concurrently and we checked for any leaked location marked by $P_a$.

As {\em per} Figure~\ref{fig:regs_attack}, in this case $P_b$ succeeds in accessing memory locations reserved by $P_a$ 
before this latter executes the \emph{cudaFree} memory releasing operation. 
Note that this behavior is different from the one observed for the global memory attack described in Section~\ref{sub:glbmem_attack}. 

We ran both $P_a$ and $P_b$ with a gridsize of $2$ blocks and a blocksize of $32$ threads.
In Table~\ref{tab:reg-attack-summary} the number of bytes of $P_a$ that are read by process $P_b$ is reported.
The analysis was conducted by varying the amount of registers declared by $P_b$ and by varying the amount of memory locations declared by $P_a$.
Results show two rounds of the experiment.
As an example, if $P_b$ reserves a 32KB register space (corresponding to 8K 32-bit registers), and $P_a$ allocates an amount of memory equal to $32MB$,
by executing the mentioned experiment twice, we obtain an information leakage of 64KB (i.e. 16K 32-bit words).
This is due to the fact that in different rounds the leakage comprises different memory locations.
The rationale is the dynamic memory management mechanism that is implemented in the GPU driver and in the CUDA Runtime.
As a consequence, this attack is even more dreadful as the adversary, 
by executing several rounds, can potentially read the whole memory segment allocated by $P_a$. \\

In order to better quantitatively evaluate this phenomenon, 
we have investigated and analyzed the relationship between the number of locations where the leakage succeeds and the number of executed rounds.
In Figure~\ref{fig:reg-rounds-kepler} results are shown with respect to a number of rounds ranging from $1000$ to $10000$.
Growth is linear in the number of rounds.
In particular, the leakage starts from $32K$ locations for $1000$ rounds  and reaches $320K$ locations leaked when the number of rounds is $10000$.
The leaked locations belong to contiguous memory areas; the distance between each location is 32 bytes.
For example, if the leaked locations start from byte $0$ and end at byte $320$, then we obtain $10$ locations: one location every $32$ bytes.
We claim that this behavior depends on the implementation and the configuration of the kernel.
A more thorough analysis will be provided in future works.

\begin{color}{\clr}
\label{writingbug} 
The results of this experiment suggest a further study 
on the possibility for a malicious process to obtain write access to the leaked locations. 
In order to investigate this vulnerability, we performed an additional set of experiments. \\
We kept the same configuration as the previous test but, 
to foster the detection of the potential unauthorized write accesses, we used a cryptographic hash function. 
Indeed, thanks to the properties of hash functions, if the malicious process succeeds in interfering with the computation of the legitimate process---for instance,  
by altering even only a single bit---, this would cause (with overwhelming probability) errors in the output of the legitimate process. \\
For $P_a$, we used a publicly-available GPU implementation of the \emph{SHA-1} hash function included into the 
\emph{SSLShader}\footnote{The source code is available at http://shader.kaist.edu/sslshader at the time of writing}:
an SSL reverse proxy transparently translating SSL sessions to TCP sessions for back-end servers.  
The encryption/decryption is performed on-the-fly through a CUDA implementation of several cryptographic algorithms. 
Actually, the codes that implements the GPU cryptograhpic algorithm, are contained in the \emph{libgpucrypto} library which can be downloaded from the same web site. 
For our experiments we used the version $0.1$ of this library. 

In this test, $P_a$ uses the GPU to compute the \emph{SHA-1} for $4096$ times on a constant plaintext of $16$KB. 
$P_a$ stores each hash in a different memory location. 
To test the integrity of GPU-computed hashes, $P_a$ also computes the \emph{SHA-1} on the CPU using the OpenSSL library 
and then compares this result with the ones computed on the GPU. \\
The malicious process $P_b$ tries to taint the computation of $P_a$ by writing a constant value into the leaked locations. 
The following PTX instruction is used:\\

\begin{Verbatim}[frame=single]
asm("mov.u32 r1,\%0;":"=r"(value));
asm("mov.u32 r2,\%0;":"=r"(value));
...
asm("mov.u32 rMAX,\%0;":"=r"(value));
\end{Verbatim}

where \emph{value} is an \emph{uint32\_t} constant value and \emph{MAX} represents the number of registers reserved by $P_b$.\\
In our test we ran $P_a$ $1000$ times and concurrently we launched the malicious process $P_b$. 
Even if in most cases we were able to read a portion of the memory reserved to $P_a$, 
the write instruction was ignored and all the hashes computed on the GPU were correct. 
In conclusion, our experiments' findings show that the register spilling vulnerability does not seem to allow interfering with the computation of the legitimate process.\\  
\end{color}

\section{Case Study: SSLShader - GPU accelerated SSL}
\label{aes-leakage-glb} 
\SetCommentSty{textit} 
\begin{algorithm}[t]
  \label{alg:aes-leak}
  
  \KwIn{\\
	$\vec{M}$: The plaintext \\
	$l$: The length of the plaintext\\
	$\vec{K}$: array of identifiers of the encryption key\\ 
  }
  \KwOut{TRUE if the attack succeeds, FALSE otherwise}
  
  \BlankLine
  \textbf{FindLeakage}($\vec{M}$,$l$,$\vec{K}$)
  \BlankLine
  \Begin{

   	s $\gets$ size of current allocable global memory on GPU\\  	
  	P $\gets$ cudamalloc(s)\\
	$j \gets 0$		\\
     \While{ $j < s$ } {
     	w $\gets$ P[j] \\
     	\uIf{$w \in \vec{K}$}{
     		cudamemset(P,0,s) /*zeroing*/\\
     		\Return{TRUE};
     	} \uElseIf{$w == M[0]$}{
     	   $i \gets 0$	\\	
     	   \While{ $i < l$ } {
     	   		\uIf{ $P[j+i]$ != $M[i]$ } {
   		     		cudamemset(P,0,s) /*zeroing*/\\
     	   			\Return{FALSE}
     	   		}
     	   		i++\\
     	   }
			cudamemset(P,0,s) /*zeroing*/\\
     		\Return{TRUE}
     	}
     	j++\\
    }
	cudamemset(P,0,s) 

    \Return{FALSE}	
  }
  \caption{The pseudo-code of the attacking process in the AES case study.}
\end{algorithm}

\begin{figure}[t]
  \centering
  \includegraphics[scale=0.6]{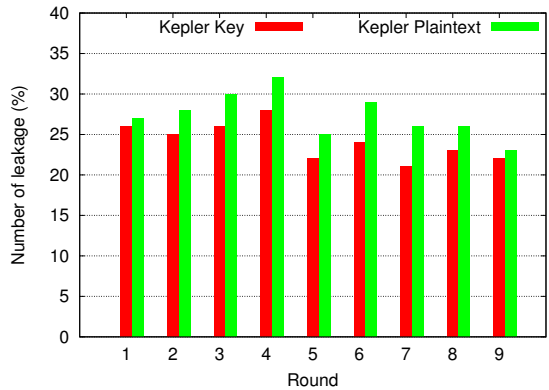}
  
  \caption{The number of leakages in the Kepler architecture. Each bin in this Figure represents the number of times that the leakage occured in $100$ runs of Algorithm \ref{alg:aes-leak}.
  We report the results for $9$ rounds of this experiments} 
  \label{fig:kepler-rounds}
\end{figure}

\begin{figure}[t]
  \centering
  \includegraphics[scale=0.6]{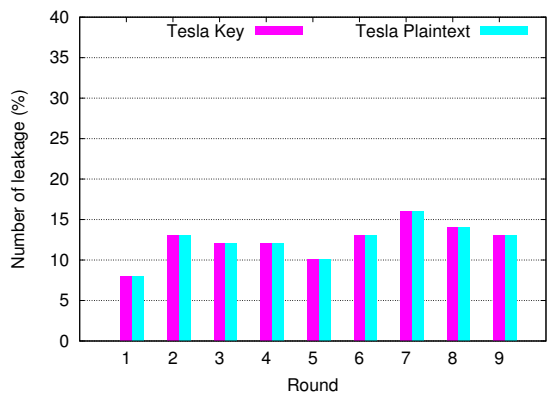}
    \caption{The number of leakages in the Fermi architecture. Each bin in this Figure represents the number of times that the leakage occured in $100$ runs of Algorithm \ref{alg:aes-leak}.
  We report the results for $9$ rounds of this experiments}
  \label{fig:tesla-rounds}
\end{figure}

In order to evaluate the impact of the global memory vulnerability in a real-world scenario, we attacked
the CUDA implementation of AES presented in \cite{Jang:2011:SCS:1972457.1972459} which is part of \emph{SSLShader}.  
The SSLShader comes with several utilities that can be used to verify the correctness of the implemented algorithms. 
To run our experiments, we modified one of these utilities. 
In particular, we changed the AES test utility to encrypt a constant plaintext using a fixed key. 
We chose a constant plaintext of $4KB$ (i.e. the first two Chapters of the \emph{Divine Comedy} written in latex) 
and a we set the $128$-bit encryption key to the juxtaposition of following words: \emph{0xdeadbeef, 0xcafed00d, 0xbaddcafe,0x8badf00d}. \\
In this experiment we assume that the GPU is shared between the adversary and the legitimate process. 
Further, we assume that the adversary can read the ciphertext. \\
The steps performed by the attacking process are described in Algorithm~\ref{alg:aes-leak}. 
We consider the attack successful in two cases:
in the first case, the adversary gets access to the \emph{whole} plaintex (line 11)---achieving \emph{plaintext leakage}; 
in the second case, the adversary obtains some words of the encryption key (line 8)---achieving \emph{key leakage}.
Even if this latter case is less dreadful than the former one, it still jeopardizes security. 
Indeed, in order to obtain the desired information, the adversary could attack the undisclosed portion of the key (e.g. via brute force, differential cryptanalysis \cite{Heys:2002:TLD:763194.763197}) and eventually decrypt the message. \\
The experiment is composed of the following steps: 
first we run an infinite loop of the CUDA AES encryption; 
in the meanwhile we run Algorithm~\ref{alg:aes-leak} $100$ times. 
In order to avoid counting a single leakage event more than once, each execution of the Algorithm~\ref{alg:aes-leak} zeroes the memory (lines $9,15,19,24$). 
We repeated this experiment $50$ times on both the Kepler and the Fermi architectures, measuring the amount of successful attacks per round. 
In order to preserve the independence across different rounds, at the end of each round we rebooted the machine. \\
For the Kepler we measured a successful attack mean equal to $30\%$ with a standard deviation equal to $0.032$.
As for the Fermi architecture, we measured a mean success rate of $12\%$ with a standard deviation of $0.03$. 
Figure~\ref{fig:kepler-rounds} details the results of $9$ randomly chosen rounds in terms of \emph{key leakage} and \emph{plaintext leakage} for Kepler. 
The \emph{plaintext leakage} is slightly more frequent than the \emph{key leakage}. 
Figure~\ref{fig:tesla-rounds} shows the results for Fermi; in this case the frequencies of the two leakages are equal. \\

\subsection{Discussion and qualitative analysis}
\label{leakagequality2} 
It is worth adding further considerations about the conditions 
that lead to an effective information leakage. 
Indeed, with our attack methodology we are able to leak only the final state of the previous GPU process.  
This limitation is due to the exclusive access granted by the driver to host threads that access the GPU; only one 
cudaContext is allowed to access the GPU at a given time. \\
However, note that in some circumstances the final state of a computation is sensitive 
(e.g. the decryption of a ciphertext, the output of a risk-analysis function or the Universal Transverse Mercator (UTM) locations of an oil well). 
In other circumstances the final state of a computation is not sensitive or even public. 
For instance, knowing the final state of an encryption process (i.e. the ciphertext) does not represent a threat.\\ 
However, in our experiments we were able to recover the original plaintext even after the encryption process ended. 
This was possible due to the fact that the plaintext and the ciphertext were stored in different locations of the global memory. 
This condition allowed us to bypass the \emph{final-state} limitation.
As such, this vulnerability depends on both the implementation and the computed function and does not hold in the general case.\\
Another important consideration about the presented case study concerns the precondition of the attack. 
In order to perform the \emph{key leakage} test we assume that the adversary knows a portion of the key 
(which is needed to perform searches in memory and detect if the leakage happened). 
However, as shown in~\cite{coldboot,zeus} it is possible to exploit the high entropy of the encryption keys to restrict the possible candidates to a reasonable number.
In fact, secure encryption keys usually have a higher entropy than other binary data in memory.

As a further technical note we found out that SSLShader \cite{Jang:2011:SCS:1972457.1972459}, 
i.e. the widespread publicly available implementation of cryptographic algorithms on GPU, 
makes use of the \emph{cudaHostAlloc} CUDA primitive. 
This primitive allocates a memory area in the host memory that is page-locked and accessible to the device (pinned memory). 
Although this can be considered more secure than using the cudaMalloc, it is fully vulnerable to information leakage as well. 
In fact, the CUDA Runtime copies the data to the GPU's global memory on behalf of the programmer when it is more convenient. 
This is important since it shows that even code implemented by ``experts'' actually shows the same deficiencies as regards security. 
This finding, together with the others reported in the paper, call for solutions to this severe vulnerability.

\smallskip
\smallskip
\smallskip
\section{Proposed Countermeasures}
\label{countermeasures}
\label{remedies}
In previous sections the main issues and vulnerabilities of Kepler and Fermi CUDA architectures have been highlighted.
This section suggests alternative approaches and countermeasure that prevent or at least dramatically limit the described information leakage attacks.

In general, from the software point of view, CUDA code writers should pay particular attention
to zeroing memory as much as possible at the end of kernel execution. 

Unfortunately, this is troublesome for a number of reasons:
\begin{itemize}
 \item Most often the programmer does not have fine control over kernel code
 (e.g. if the kernel is the outcome of high-level programming environments such as JavaCL, Jcuda \cite{jcuda}, etc);
 \item The kernel programmer usually aims at writing the fastest possible code
 without devoting time to address security/isolation issues that might hamper performance.
\end{itemize}

As such, we believe that the best results can be obtained if security-enhancements are performed at the driver/hardware level.
From the CUDA Platform/Hardware point of view:
\begin{itemize}
 \item Finer MMU memory protection 
 must be put in place to prevent concurrent kernels from reading other kernel's memory; 
 \item CUDA should also allow the OS to collect and access forensics information (this hypothesis requires the host is trusted).
\end{itemize}

In the following, for each of the discovered vulnerabilities, we provide related mitigation countermeasures.
\subsection{Shared Memory}
As for  the shared memory leakage shown in Section \ref{sub:shmem_attack}, the proposed fix makes use of a memory-zeroing mechanism.
As already pointed out in Section~\ref{sub:shmem_attack} the shared memory attack is ineffective once the host process terminates.
The vulnerability window goes from kernel completion to host process completion.
As a consequence, the memory-zeroing operation is better executed inside the kernel.
In our opinion, this is a sensitive solution since shared memory is an on-chip area that cannot be directly addressed or copied by the host thread.
As such, it is not possible to make use of it from outside kernel functions.

\label{zeroingvector} 
In order to measure the overhead that an in-kernel memory-zeroing approach would have on a real GPU,
we developed and instrumented a very simple CUDA code (addition of two vectors).
Two kernel functions, $K_1$ and $K_2$ were developed:
$K_1$ receives as input two randomly-initialized vectors $A,B$;
$K_1$ sums the two vectors and stores the result in vector $C$;
$K_2$ is the same as $K_1$ but in addition it ``zeroes'' the shared memory area by overwriting it
with the value read from $A[0]$\footnote{we did not actually ``zero'' the memory using the value $0$
to prevent the compiler from performing optimization that would have affected the result.}. 
We measured the execution time difference between $K_1$ and $K_2$ by varying the vectors' size,
as this experiment was meant to evaluate the scalability of the zeroing operation. 
In particular, for $K_1$ and $K_2$, we performed this experiment accessing an increasing number of locations up to the maximum available shared memory.
Such value depends on the GPU capability (see Table \ref{tab:testbed}) and corresponds to 672KB for the Tesla C2050 and 32KB for the GT640.
We noticed that the introduced overhead was constant and not affected by the number of memory accesses.
In particular we 
measured a mean overhead of $1.66$ \emph{ms} on the Kepler
and $0.27$ \emph{ms} on the Tesla card.
We concluded that the proposed fix can be applied without noticeably affecting GPU performance.

\subsection{Global Memory}

\begin{figure}
	\centering
	\includegraphics[scale=0.35,angle=-90]{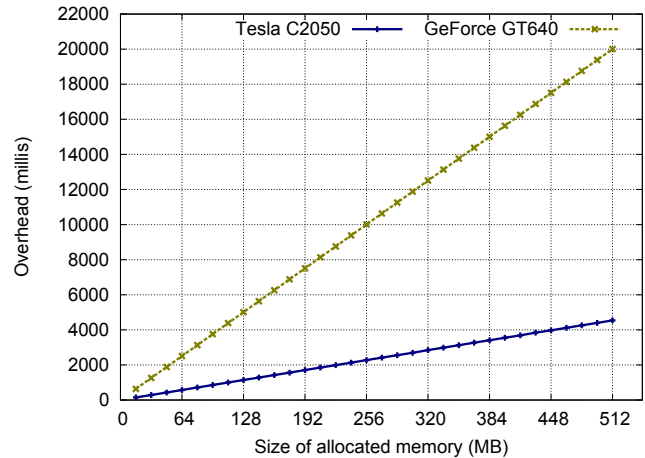}
	\caption{Overhead introduced by the proposed countermeasure for the global memory leak}
	\label{fig:gbl-zero}
\end{figure}

As described in Section\ref{sub:glbmem_attack},
accessing global memory through CUDA primitives can cause an information leakage.
The natural fix would consist in zeroing memory before it is given to the requesting process.
This way, when information is deleted the malicious process/party is not able to access such information.
This approach should naturally be implemented inside the CUDA Runtime.
To assess 
the impact of the overhead introduced by this solution, we measured the overhead
that the same zeroing operation imposes to traditional memory allocation.
CUDA Runtime function \emph{cudaMemset} was used for zeroing memory content.
An incremental size buffer was tested in the experiments.
In our tests, size ranged from 16MB to 512MB, in steps of 16MB.
We then measured the overhead induced by the additional zeroing operations.
To achieve this goal, we instrumented the source code with the \emph{EventManagement} Runtime library function.
Through these primitives we were able to compute the time elapsed between two events in milliseconds
with a resolution of around 0.5 microseconds.\\
As shown in Figure~\ref{fig:gbl-zero}, the introduced overhead is not negligible.
On both Tesla and Kepler platforms the induced overhead is linearly proportional to allocated buffer size.
However, Tesla's line steepness is much lower than the Kepler counterpart.
The reason is that the two GPUs feature a much different number of multiprocessors
(14 for Tesla vs. 2 for the Kepler). As such the level of achievable parallelism is quite different.\\

\begin{color}{\clr}
\label{zeroing}
As regards the implemented memory-zeroing approach, threads run in parallel, each one zeroing its serial memory area. 
This accounts for the measured overhead. However, since the data chunk has a very limited size, the overhead is---in absolute values---very small
($1.66$ \emph{ms} Kepler $0.27$ \emph{ms} TeslaFermi). 
It is worth noting that a low-level hardware approach would be surely faster.
However, in general zeroing does worsen performance in GPU \cite{zeroingoverhead}, 
as these techniques force additional memory copies between host and device memory.
In addition, we only have implementation details on global memory that is actually implemented on commodity GDDRx memory, i.e. as standard host memory.
Introducing an additional mechanism to perform smart memory zeroing would require an overall redesign of the GDDR approach and as such it will most probably increase RAM cost. 
Hardware-based fast zeroing would probably be the most feasible and conevenient solution.
However, inner details about low-level memory implementation for CUDA cards are only known by Nvidia.

Pertaining to the selective deletion of sensitive data, 
selectively zeroing specific memory areas is potentially feasible
and would potentially reduce unnecessary memory transfers between GPU and CPU,
since most data would not have to be transferred again.
A ``smart'' solution would probably be the addition of CUDA language extensions (source code tags) 
to mark the variables/memory areas that have to be zeroed since containing sensitive data.
On the one hand, this would require language/compiler modifications while, on the other hand, it would save some costly data transfers. 
However, this approach implies some caveats,
as there is the risk of pointing the adversary exactly to the memory and registers where sensitive data is.
Further, such sensitive data when in transit between CPU and GPU crosses various memory areas that are still potentially accessible.
As such, for performance sake, sensitive areas should be as contiguous as possible.
\end{color}

\subsection{Registers}

\begin{color}{\clr}
Register allocation is handled at lower level in the software stack, 
hence the leak is probably due to an implementation bug regarding the memory isolation module.
Therefore, fixing this leakage at the application level is quite difficult.
A much simpler workaround would be to implement the fix at the GPU driver level.
Unfortunately, given the closed-source nature of the driver, at present only NVIDIA can provide a fix for this issue.
In particular, the driver should preserve the following properties: 
first, the registers should not spill to locations in global memory that are still reserved for host-threads;
second, the locations of the spilled registers must be reset to zero when they are released.
\end{color}

\section{Related Work}
\label{related}
Information leakage is a serious problem that has been addressed in a variety of different scenarios.\\*

Many side channel attacks on cache-timing have been proposed in the literature (\cite{shamir-aes}, \cite{des-sidechan},
\cite{Menichelli:2008:HSA:1439183.1439192}).
Such attacks are relevant as they exploit the vulnerabilities of the underlying HW/SW architecture.
\begin{color}{\clr}
\label{cpuleakage} 
In particular, Shamir et al. \cite{shamir-aes} managed to infer information about the internal state of a cipher by 
exploiting time-based side channels on the x86 CPU caches.
We have to say that CPU caches are quite different from GPU shared memory considered in the present paper \cite{Cope:2010:PCG:1749402.1749493}. 
Indeed, contrary to CPU cache memory, GPU shared memory is addressable by the programmer \cite{Xu:2012:OSM:2376362.2376726}. 
In the GPU environment, the adversary does not need to exploit time side-channels to infer information about the shared memory since the adversary can just read it. 
Modern GPUs are equipped with a hierarchical cache system.
However, timing facilities in the CPU are much more precise than the coarse grained timing hardware currently available on GPUs.
As such, exploiting time-based side channel of GPU cache memories would require a specific investigation effort that 
is out of the scope of present work.
\end{color}

In \cite{6296708} the authors analyze a category of side-channel attacks known as profiled cache-timing 
attacks, and develop a methodology that allows an adversary (capable of a limited number of side-channel measurements) 
to choose the best attack strategy. 
In \cite{Kang93apump}, the authors propose a HW solution to minimize the capacity of covert timing channel across
different levels of trust in Multilevel Security Systems. 
However, in \cite{TIFS-pump} the authors devise a technique to compromise the security provided by these hardware components.
Note that in this work we do not resort to timing side-channels as the tests we performed indicated that reliable 
timing is quite hard to achieve on the GPU platforms used for the experiments.\\
\begin{color}{\clr}
Evidence exists that security and isolation are not considered as important as performance. 
In fact, the trend towards increased resource sharing among cores is represented by Gupta et al. \cite{Gupta:2010:ECB:1934902.1934979} that encourage 
resource sharing inside future multicores for performance, fault tolerance and customized processing. 
The vision suggests reducing isolation among cores for the sake of performance and reliability.
However, this opens up new interesting information leakage opportunities.
Oz et al. \cite{Oz:2012:TVP:2350366.2350516} propose and evaluate a new reliability metric called the Thread Vulnerability Factor (TVF) that depends on its code but also on the codes of sibling threads.
Their evaluation shows that TVF values tend to increase as the number of cores increases, which means the system becomes more vulnerable as the core count rises. 

\end{color}
A preliminary work by Barenghi et al. \cite{springerlink:10.1007/978-3-642-21040-29} investigated side channel attacks to GPUs
using both power consumption and electromagnetic (EM) radiations.
The proposed approach can be useful for GPU manufacturers to protect data against physical attacks.
However, for our attacks to work, the adversary does not need either physical access to the machine
or root privileges.
Protecting from an adversary with super-user (or root) administration privileges is very difficult, as shown in \cite{securecloud}, and it is out of the scope of present work. 

For what concerns malware-related leaks, Trusted Platform Modules (TPM) \cite{tifstpm, tifstpm2} and remote attestation can provide an acceptable
level of security by leveraging secure boot.
However, vulnerabilities can also stem from perfectly ``legal'' code 
that accesses other parties' data exploiting vulnerabilities inherently tied to the platform.
Moreover, malicious hardware such as trojan circuitry can bypass software TPM mechanisms and access 
sensitive information over the bus. 
In \cite{Das:2010:DIL:1870926.1871135} an architecture is proposed  based on an external guardian 
core that is required to approve each memory request. 
Even though the induced performance overhead is too high (60\%) and even though we do not consider
malicious hardware in this work, the work in \cite{Das:2010:DIL:1870926.1871135} is particularly interesting, 
as actual cooperation from the hardware (i.e. from its manufacturers) would be beneficial for information leakage detection and prevention.

Unfortunately, due to Companies commercial strategy to hide implementation details from competitors, 
manufacturers are reluctant on publishing the internals of their solutions. 
In fact, documentation is mostly generic, marketing-oriented, and incomplete. 
This fact hinders the analysis of the information leakage problem in the context of GPU.
As such, in the literature most of the available architectural information over existing hardware is due to black-box analysis.
In particular \cite{demystifying} developed a microbenchmark suite to measure architectural characteristics of CUDA GPUs.
The analysis showed various undisclosed characteristics of the processing elements and the memory hierarchies
and exposed undocumented features that impact both program performance and program correctness.
CUBAR \cite{cubar} used a similar approach to discover some of the undisclosed CUDA details.
In particular, CUBAR showed that CUDA features a Harvard architecture on a Von Neumann unified memory.
Further, since the closed-source driver leverages (the deprecated) security through obscurity paradigm,
inferring information from PCIe bus \cite{cudadevguide} is possible as partially shown in \cite{gdev}.\\
The main contribution on GPU security in the literature are mainly related to the integrity of the platform, 
and to the exploitation of driver vulnerabilities.
GPU thread synchronization issues are introduced and discussed in \cite{5537722}
whereas reliability of multicore computing is discussed in \cite{Shye:2009:PSA:1550410.1550669}.
Such analysis are aimed towards correctness, reliability and performance, 
whereas in this work we focus on actual GPU thread  behavior and related consequences on data access.
Further, vulnerabilities have been discovered in the past in the NVIDIA GPU driver \cite{nvidiavuln1, nvidiavuln2}.
The device driver is a key component of the CUDA system and has kernel level access (via the NVIDIA kernel module).
As such, vulnerabilities in the CUDA system can have nasty effects on the whole system 
and can lead to even further information leakage, due to root access capabilities.\\*
A limitation of the current GPU architecture is related to the fact that the OS is
completely excluded from the management of computations that have to be performed on the device. 
The first attempts to overcome the limits of present GPU platforms aim at giving the OS kernel
the ability to control the marshalling of the GPU tasks \cite{timegraph,ptask}.
As a matter of fact, the GPU can be seen as an independent computing system where the OS role is played by the GPU device driver;
as a consequence, host-based memory protection mechanism are actually ineffective to protect GPU memory.

\balance

\section{Conclusion and Future Work}
\label{conclusion}

This work provides a relevant new contribution to the security of the increasingly successful GPGPU computing field.
in the CUDA architecture affecting shared memory, global memory, and registers, that can actually lead to 
information leakage. 
Further, we propose and discuss some fixes 
to tackle the highlighted vulnerabilities. 
As for future work, we are currently investigating GPGPU information leakage issues within Windows OSes. 

\section*{Acknowledgment}
The authors would like to thank the anonymous reviewers for their helpful comments.
This work has been partly supported by an NVIDIA Academic Research donation.
Roberto Di Pietro has been partly supported by a Chair of Excellence from University Carlos
III, Madrid.
\balance
\bibliographystyle{IEEEtran}
\bibliography{complessivo2}

\begin{thebibliography}{10}
\providecommand{\url}[1]{#1}
\csname url@samestyle\endcsname
\providecommand{\newblock}{\relax}
\providecommand{\bibinfo}[2]{#2}
\providecommand{\BIBentrySTDinterwordspacing}{\spaceskip=0pt\relax}
\providecommand{\BIBentryALTinterwordstretchfactor}{4}
\providecommand{\BIBentryALTinterwordspacing}{\spaceskip=\fontdimen2\font plus
\BIBentryALTinterwordstretchfactor\fontdimen3\font minus
  \fontdimen4\font\relax}
\providecommand{\BIBforeignlanguage}[2]{{%
\expandafter\ifx\csname l@#1\endcsname\relax
\typeout{** WARNING: IEEEtran.bst: No hyphenation pattern has been}%
\typeout{** loaded for the language `#1'. Using the pattern for}%
\typeout{** the default language instead.}%
\else
\language=\csname l@#1\endcsname
\fi
#2}}
\providecommand{\BIBdecl}{\relax}
\BIBdecl

\bibitem{titan}
OLCF, \url{http://www.olcf.ornl.gov/titan/}, 2012.

\bibitem{computationalfinancegpu}
A.~Gaikwad and I.~M. Toke, ``Parallel iterative linear solvers on {GPU}: A
  financial engineering case,'' in \emph{Proceedings of the 2010 18th Euromicro
  Conference on Parallel, Distributed and Network-based Processing}, ser. PDP
  '10.\hskip 1em plus 0.5em minus 0.4em\relax Washington, DC, USA: IEEE
  Computer Society, 2010, pp. 607--614.

\bibitem{aes-gpu}
A.~Di~Biagio, A.~Barenghi, G.~Agosta, and G.~Pelosi, ``Design of a parallel
  {AES} for graphics hardware using the {CUDA} framework,'' in \emph{Parallel
  Distributed Processing, 2009. IPDPS 2009. IEEE International Symp. on}, may
  2009, pp. 1 --8.

\bibitem{aescuda}
C.~Mei, H.~Jiang, and J.~Jenness, ``{CUDA}-based {AES} parallelization with
  fine-tuned {GPU} memory utilization,'' in \emph{Parallel Distributed
  Processing, Workshops and Phd Forum (IPDPSW), 2010 IEEE International Symp.
  on}, april 2010, pp. 1--7.

\bibitem{aes-xts}
M.~Alomari and K.~Samsudin, ``A framework for {GPU}-accelerated {AES-XTS}
  encryption in mobile devices,'' in \emph{TENCON 2011 - 2011 IEEE Region 10
  Conference}, nov. 2011, pp. 144 --148.

\bibitem{symmetriccudageneric}
N.~Nishikawa, K.~Iwai, and T.~Kurokawa, ``High-performance symmetric block
  ciphers on {CUDA},'' in \emph{Networking and Computing (ICNC), 2011 Second
  International Conference on}, 30 2011-dec. 2 2011, pp. 221 --227.

\bibitem{designingGPUencryption}
G.~Barlas, A.~Hassan, and Y.~Al~Jundi, ``An analytical approach to the design
  of parallel block cipher encryption/decryption: A {CPU/GPU} case study,'' in
  \emph{Parallel, Distributed and Network-Based Processing (PDP), 2011 19th
  Euromicro International Conference on}, feb. 2011, pp. 247 --251.

\bibitem{aescudaanalysis}
K.~Iwai, T.~Kurokawa, and N.~Nisikawa, ``{AES} encryption implementation on
  {CUDA} {GPU} and its analysis,'' in \emph{Networking and Computing (ICNC),
  2010 First International Conference on}, nov. 2010, pp. 209 --214.

\bibitem{gpuparallcrypto}
R.~Detomini, R.~Lobato, R.~Spolon, and M.~Cavenaghi, ``Using {GPU} to exploit
  parallelism on cryptography,'' in \emph{Information Systems and Technologies
  (CISTI), 2011 6th Iberian Conference on}, june 2011, pp. 1 --6.

\bibitem{Georgescu:2011:GAC:2082156.2082161}
S.~Georgescu and P.~Chow, ``Gpu accelerated cae using open solvers and the
  cloud,'' \emph{SIGARCH Comput. Archit. News}, vol.~39, no.~4, pp. 14--19, dec
  2011.

\bibitem{actualgpuinthecloud}
Softlayer, ``Softlayer adds high-performance computing to global platform,''
  \url{http://www.softlayer.com/dedicated-servers/high-performance-computing},
  2012.

\bibitem{vgpuvirtualization}
Zillians, ``{VGPU} {GPU} virtualization,''
  \url{http://www.zillians.com/products/vgpu-gpu-virtualization/}, 2012.

\bibitem{vgxhypervisor}
Citrix, ``{NVIDIA} {VGX} hypervisor,''
  \url{http://www.nvidia.com/object/vgx-hypervisor.html}, 2012.

\bibitem{cudadma}
M.~Bauer, H.~Cook, and B.~Khailany, ``{CudaDMA}: optimizing gpu memory
  bandwidth via warp specialization,'' in \emph{Proceedings of 2011
  International Conference for High Performance Computing, Networking, Storage
  and Analysis}, ser. SC '11.\hskip 1em plus 0.5em minus 0.4em\relax New York,
  NY, USA: ACM, 2011, pp. 12:1--12:11.

\bibitem{nvidiavuln1}
M.~Larabel, ``{NVIDIA} {L}inux driver hack gives you root access,''
  \url{http://www.phoronix.com/scan.php?page=news_item&px=MTE1MTk}, 2012.

\bibitem{nvidiavuln2}
------, ``{NVIDIA} 295.40 closes high-risk security flaw,''
  \url{http://www.phoronix.com/scan.php?page=news_item&px=MTA4NTk}, 2011.

\bibitem{gdev}
S.~Kato, ``Gdev,'' \url{https://github.com/shinpei0208/gdev}, 2012.

\bibitem{kgpu}
W.~Sun, R.~Ricci, and M.~L. Curry, ``{KGPU} harnessing {GPU} power in {L}inux
  kernel,'' \url{http://code.google.com/p/kgpu}, 2012.

\bibitem{linuxinthecloud}
B.~Butler, ``Linux distributors duke it out in cloud os market,''
  \url{http://www.networkworld.com/news/2012/080812-linux-cloud-261495.html},
  2012.

\bibitem{gpgpu}
D.~Luebke, M.~Harris, J.~Kr\"{u}ger, T.~Purcell, N.~Govindaraju, I.~Buck,
  C.~Woolley, and A.~Lefohn, ``{GPGPU}: general purpose computation on graphics
  hardware,'' in \emph{ACM SIGGRAPH 2004 Course Notes}, ser. SIGGRAPH
  '04.\hskip 1em plus 0.5em minus 0.4em\relax New York, NY, USA: ACM, 2004.

\bibitem{gpgpu2}
Y.~Yang, P.~Xiang, J.~Kong, M.~Mantor, and H.~Zhou, ``A unified optimizing
  compiler framework for different gpgpu architectures,'' \emph{ACM Trans.
  Archit. Code Optim.}, vol.~9, no.~2, pp. 9:1--9:33, jun 2012.

\bibitem{jcuda}
Y.~Yan, M.~Grossman, and V.~Sarkar, ``Jcuda: A programmer-friendly interface
  for accelerating java programs with cuda,'' in \emph{Proceedings of the 15th
  International Euro-Par Conference on Parallel Processing}, ser. Euro-Par
  '09.\hskip 1em plus 0.5em minus 0.4em\relax Berlin, Heidelberg:
  Springer-Verlag, 2009, pp. 887--899.

\bibitem{cudadevguide}
{NVIDIA}, ``{CUDA} 4.2 developers guide,''
  \url{http://developer.download.nvidia.com/compute/DevZone/docs/html/doc/CUDA_C_Programming_Guide.pdf},
  2012.

\bibitem{cudacs}
F.~Lombardi and R.~Di~Pietro, ``{CUDACS}: securing the cloud with cuda-enabled
  secure virtualization,'' in \emph{Proceedings of the 12th international
  conference on Information and communications security}, ser. ICICS'10.\hskip
  1em plus 0.5em minus 0.4em\relax Berlin, Heidelberg: Springer-Verlag, 2010,
  pp. 92--106.

\bibitem{coalescing}
B.~Jang, D.~Schaa, P.~Mistry, and D.~Kaeli, ``Exploiting memory access patterns
  to improve memory performance in data-parallel architectures,'' \emph{IEEE
  Trans. Parallel Distrib. Syst.}, vol.~22, no.~1, pp. 105--118, jan 2011.

\bibitem{zeroizzazione}
D.~Evans, P.~Bond, and A.~Bement, ``Fips pub 140-2 security requirements for
  cryptographic modules,'' 1994.

\bibitem{zeroizzazione2}
H.~C.~A. van Tilborg and S.~Jajodia, Eds., \emph{Encyclopedia of Cryptography
  and Security, 2nd Ed}.\hskip 1em plus 0.5em minus 0.4em\relax Springer, 2011.

\bibitem{gmac}
I.~Gelado, ``Gmac: Global memory for accelerators,''
  \url{http://developer.download.nvidia.com/compute/DevZone/docs/html/doc/CUDA_C_Programming_Guide.pdf},
  2012.

\bibitem{securitythroughobscurity}
R.~T. Mercuri and P.~G. Neumann, ``Security by obscurity,'' \emph{Commun. ACM},
  vol.~46, no.~11, pp. 160--166, nov 2003.

\bibitem{zeroingoverhead}
X.~Yang, S.~M. Blackburn, D.~Frampton, J.~B. Sartor, and K.~S. McKinley, ``Why
  nothing matters: the impact of zeroing,'' \emph{SIGPLAN Not.}, vol.~46,
  no.~10, pp. 307--324, oct 2011.

\bibitem{Jang:2011:SCS:1972457.1972459}
K.~Jang, S.~Han, S.~Han, S.~Moon, and K.~Park, ``Sslshader: cheap ssl
  acceleration with commodity processors,'' in \emph{Proceedings of the 8th
  USENIX conference on Networked systems design and implementation}, ser. NSDI
  11.\hskip 1em plus 0.5em minus 0.4em\relax Berkeley, CA, USA: USENIX
  Association, 2011, pp. 1--1.

\bibitem{Heys:2002:TLD:763194.763197}
H.~M. Heys, ``A tutorial on linear and differential cryptanalysis,''
  \emph{Cryptologia}, vol.~26, no.~3, pp. 189--221, jul 2002.

\bibitem{coldboot}
J.~A. Halderman, S.~D. Schoen, N.~Heninger, W.~Clarkson, W.~Paul, J.~A. Cal,
  A.~J. Feldman, and E.~W. Felten, ``Least we remember: Cold boot attacks on
  encryption keys,'' in \emph{In USENIX Security Symposium}, 2008.

\bibitem{zeus}
M.~Riccardi, R.~{Di Pietro}, M.~Palanques, and J.~A. Vila, ``Titans' revenge:
  Detecting zeus via its own flaws,'' \emph{Elsevier Computer Networks}, in
  press, 2012.

\bibitem{shamir-aes}
D.~Osvik, A.~Shamir, and E.~Tromer, ``Cache attacks and countermeasures: The
  case of {AES},'' in \emph{Topics in Cryptology – CT-RSA 2006}, ser. LNCS,
  D.~Pointcheval, Ed.\hskip 1em plus 0.5em minus 0.4em\relax Springer Berlin
  Heidelberg, 2006, vol. 3860, pp. 1--20.

\bibitem{des-sidechan}
Y.~Tsunoo, T.~Saito, T.~Suzaki, and M.~Shigeri, ``Cryptanalysis of des
  implemented on computers with cache,'' in \emph{Proc. of CHES 2003, Springer
  LNCS}.\hskip 1em plus 0.5em minus 0.4em\relax Springer-Verlag, 2003, pp.
  62--76.

\bibitem{Menichelli:2008:HSA:1439183.1439192}
F.~Menichelli, R.~Menicocci, M.~Olivieri, and A.~Trifiletti, ``High-level
  side-channel attack modeling and simulation for security-critical systems on
  chips,'' \emph{IEEE Trans. Dependable Secur. Comput.}, vol.~5, no.~3, pp.
  164--176, jul 2008.

\bibitem{Cope:2010:PCG:1749402.1749493}
B.~Cope, P.~Y.~K. Cheung, W.~Luk, and L.~Howes, ``Performance comparison of
  graphics processors to reconfigurable logic: A case study,'' \emph{IEEE
  Trans. Comput.}, vol.~59, no.~4, pp. 433--448, 2010.

\bibitem{Xu:2012:OSM:2376362.2376726}
\BIBentryALTinterwordspacing
W.~Xu, H.~Zhang, S.~Jiao, D.~Wang, F.~Song, and Z.~Liu, ``Optimizing sparse
  matrix vector multiplication using cache blocking method on fermi gpu,'' in
  \emph{Proceedings of the 2012 13th ACIS International Conference on Software
  Engineering, Artificial Intelligence, Networking and Parallel/Distributed
  Computing}, ser. SNPD '12.\hskip 1em plus 0.5em minus 0.4em\relax Washington,
  DC, USA: IEEE Computer Society, 2012, pp. 231--235. [Online]. Available:
  \url{http://dx.doi.org/10.1109/SNPD.2012.20}
\BIBentrySTDinterwordspacing

\bibitem{6296708}
C.~Rebeiro and D.~Mukhopadhay, ``Boosting profiled cache timing attacks with
  apriori analysis,'' \emph{Information Forensics and Security, IEEE
  Transactions on}, vol.~PP, no.~99, p.~1, 2012.

\bibitem{Kang93apump}
M.~H. Kang and I.~S. Moskowitz, ``A pump for rapid, reliable, secure
  communication,'' in \emph{Proc. ACM Conf. Computer and Communication
  Security}, 1993, pp. 119--129.

\bibitem{TIFS-pump}
S.~Gorantla, S.~Kadloor, N.~Kiyavash, T.~Coleman, I.~Moskowitz, and M.~Kang,
  ``Characterizing the efficacy of the {NRL} network pump in mitigating covert
  timing channels,'' \emph{Information Forensics and Security, IEEE
  Transactions on}, vol.~7, no.~1, pp. 64 --75, february 2012.

\bibitem{Gupta:2010:ECB:1934902.1934979}
S.~Gupta, S.~Feng, A.~Ansari, and S.~Mahlke, ``Erasing core boundaries for
  robust and configurable performance,'' in \emph{Microarchitecture (MICRO),
  2010 43rd Annual IEEE/ACM International Symposium on}, dec. 2010, pp. 325
  --336.

\bibitem{Oz:2012:TVP:2350366.2350516}
I.~Oz, H.~R. Topcuoglu, M.~Kandemir, and O.~Tosun, ``Thread vulnerability in
  parallel applications,'' \emph{J. Parallel Distrib. Comput.}, vol.~72,
  no.~10, pp. 1171--1185, oct 2012.

\bibitem{springerlink:10.1007/978-3-642-21040-29}
A.~Barenghi, G.~Pelosi, and Y.~Teglia, ``Information leakage discovery
  techniques to enhance secure chip design,'' in \emph{Information Security
  Theory and Practice. Security and Privacy of Mobile Devices in Wireless
  Communication}, ser. Lecture Notes in Computer Science, C.~Ardagna and
  J.~Zhou, Eds.\hskip 1em plus 0.5em minus 0.4em\relax Springer Berlin /
  Heidelberg, 2011, vol. 6633, pp. 128--143.

\bibitem{securecloud}
F.~Lombardi and R.~Di~Pietro, ``Secure virtualization for cloud computing,''
  \emph{J. Netw. Comput. Appl.}, vol.~34, no.~4, pp. 1113--1122, jul 2011.

\bibitem{tifstpm}
R.~Fink, A.~Sherman, and R.~Carback, ``{TPM} meets {DRE}: Reducing the trust
  base for electronic voting using trusted platform modules,''
  \emph{Information Forensics and Security, IEEE Transactions on}, vol.~4,
  no.~4, pp. 628 --637, dec. 2009.

\bibitem{tifstpm2}
A.~Kanuparthi, M.~Zahran, and R.~Karri, ``Architecture support for dynamic
  integrity checking,'' \emph{Information Forensics and Security, IEEE
  Transactions on}, vol.~7, no.~1, pp. 321 --332, feb. 2012.

\bibitem{Das:2010:DIL:1870926.1871135}
A.~Das, G.~Memik, J.~Zambreno, and A.~Choudhary, ``Detecting/preventing
  information leakage on the memory bus due to malicious hardware,'' in
  \emph{Proceedings of the Conference on Design, Automation and Test in
  Europe}, ser. DATE '10.\hskip 1em plus 0.5em minus 0.4em\relax European
  Design and Automation Association, 2010, pp. 861--866.

\bibitem{demystifying}
H.~Wong, M.-M. Papadopoulou, M.~Sadooghi-Alvandi, and A.~Moshovos,
  ``Demystifying gpu microarchitecture through microbenchmarking,'' in
  \emph{Performance Analysis of Systems Software (ISPASS), 2010 IEEE
  International Symposium on}, march 2010, pp. 235 --246.

\bibitem{cubar}
N.~Black and J.~Rodzik, ``My other computer is your {GPU}: System-centric
  {CUDA} threat modeling with {CUBAR},''
  \url{http://dank.qemfd.net/dankwiki/images/d/d2/Cubar2010.pdf}, 2010.

\bibitem{5537722}
W.~chun Feng and S.~Xiao, ``To {GPU} synchronize or not {GPU} synchronize?'' in
  \emph{Circuits and Systems (ISCAS), Proceedings of 2010 IEEE International
  Symposium on}, 30 2010-june 2 2010, pp. 3801 --3804.

\bibitem{Shye:2009:PSA:1550410.1550669}
A.~Shye, J.~Blomstedt, T.~Moseley, V.~J. Reddi, and D.~A. Connors, ``Plr: A
  software approach to transient fault tolerance for multicore architectures,''
  \emph{IEEE Trans. Dependable Secur. Comput.}, vol.~6, no.~2, pp. 135--148,
  apr 2009.

\bibitem{timegraph}
S.~Kato, K.~Lakshmanan, R.~Rajkumar, and Y.~Ishikawa, ``Timegraph: {GPU}
  scheduling for real-time multi-tasking environments,'' in \emph{Proceedings
  of the 2011 USENIX conference on USENIX annual technical conference}, ser.
  USENIXATC'11.\hskip 1em plus 0.5em minus 0.4em\relax Berkeley, CA, USA:
  USENIX Association, 2011, pp. 2--2.

\bibitem{ptask}
C.~J. Rossbach, J.~Currey, M.~Silberstein, B.~Ray, and E.~Witchel, ``{PTask}:
  operating system abstractions to manage {GPUs} as compute devices,'' in
  \emph{Proceedings of the Twenty-Third ACM Symposium on Operating Systems
  Principles}, ser. SOSP '11.\hskip 1em plus 0.5em minus 0.4em\relax New York,
  NY, USA: ACM, 2011, pp. 233--248.

\end{thebibliography}
\balance
\end{document}